\documentclass{pasj01}

\usepackage{natbib}
 
\Received{$\langle$reception date$\rangle$}
\Accepted{$\langle$acception date$\rangle$}
\Published{$\langle$publication date$\rangle$}

\begin{document}

\title{ALMA observations of the dense and shocked gas 
in the nuclear region of NGC~4038 (Antennae galaxies)}
\author{
Junko \textsc{Ueda}\altaffilmark{1,2,*}, 
Yoshimasa \textsc{Watanabe}\altaffilmark{3}, 
Daisuke \textsc{Iono}\altaffilmark{2,4}, 
David J. \textsc{Wilner}\altaffilmark{1},
Giovanni G. \textsc{Fazio}\altaffilmark{1},
Satoshi \textsc{Ohashi}\altaffilmark{2,5},
Ryohei \textsc{Kawabe}\altaffilmark{2,4,5},
Toshiki \textsc{Saito}\altaffilmark{2,5}, 
Shinya \textsc{Komugi}\altaffilmark{6}
}
\altaffiltext{1}{Harvard-Smithsonian Center for Astrophysics, 60 Garden Street, Cambridge, MA 02138, USA}
\altaffiltext{2}{National Astronomical Observatory of Japan, 2-21-1 Osawa, Mitaka,Tokyo, 181-8588, Japan}
\altaffiltext{3}{Department of Physics, The University of Tokyo, 7-3-1 Hongo, Bunkyo-ku, Tokyo, 113-0033, Japan}
\altaffiltext{4}{The Graduate University for Advanced Studies (SOKENDAI), 2-21-1 Osawa, Mitaka, Tokyo 181-8588, Japan}
\altaffiltext{5}{Department of Astronomy, The University of Tokyo, 7-3-1 Hongo, Bunkyo-ku, Tokyo 133-0033, Japan}
\altaffiltext{6}{Division of Liberal Arts, Kogakuin University, 2665-1, Hachioji, Tokyo 192-0015, Japan}
\email{junko.ueda@cfa.harvard.edu}


\maketitle

\begin{abstract}
We present 1\arcsec ($<$100~pc) resolution maps of millimeter emission 
from five molecules--CN, HCN, HCO$^{+}$, CH$_{3}$OH, and HNCO--obtained towards NGC~4038, 
which is the northern galaxy of the mid-stage merger, Antennae galaxies, 
with the Atacama Large Millimeter/submillimeter Array.  
Three molecules (CN, CH$_{3}$OH, and HNCO) were detected 
for the first time in the nuclear region of NGC~4038.  
High-resolution mapping reveals a systematic difference 
in distributions of different molecular species and continuum emission.  
Active star forming regions identified by the 3~mm and 850~$\mu$m continuum emission 
are offset from the gas-rich region associated with the HCN~(1--0) and CO~(3--2) peaks.  
The CN~(1--0)/HCN~(1--0) line ratios 
are enhanced (CN/HCN $\simeq$ 0.8--1.2) in the star forming regions, 
suggesting that the regions are photon dominated.  
The large molecular gas mass (10$^{8}~M_{\solar}$) within a 0\farcs6 ($\sim$60~pc) radius of the CO~(3--2) peak 
and a high dense gas fraction ($>$20~\%) suggested by the HCN~(1--0)/CO~(3--2) line ratio 
may signify a future burst of intense star formation there.  
The shocked gas traced in the CH$_{3}$OH and HNCO emission indicates sub-kpc scale molecular shocks.  
We suggest that the molecular shocks may be driven by collisions 
between inflowing gas and the central massive molecular complex.  
\end{abstract}

\section{Introduction}
Dynamical interactions and mergers of two comparable mass gas-rich galaxies 
can result in galaxies bright in IR luminosity (e.g., ultra/luminous infrared galaxies (U/LIRGs)).  
While recent numerical studies have shown that disk wide starbursts 
produced by mass fragmentation and turbulent motion across the disk 
can contribute to the overall star formation activity in colliding systems \citep{Teyssier10}, 
it has been long predicted that the ubiquitous presence of large-scale gas inflows 
leads to central gas accumulation and subsequent nuclear starburst activity \citep{Mihos96}.  
The inflowing gas could collide with gas component associated with the nuclear disk/ring, 
causing molecular shocks.  
Although evidence for gas inflows on galactic scales has been discovered 
in neutral gas \citep[e.g.,][]{Hibbard96, Iono05} and ionized gas \citep[e.g.,][]{Rampazzo05, Kewley10}, 
there have been few observational confirmations of gas inflows on $\leq$1~kpc scale.

The Antennae galaxies (NGC~4038/9) are the nearest 
\citep[22~Mpc (1\arcsec = 107~pc);][]{Schweizer08} mid-stage merger.  
It is well known that active star formation is ongoing 
in the off-nuclear region, called the ``overlap region" \citep[e.g.,][]{Wang04}, 
and that the two nuclear regions have lower star formation rates (SFRs) than the overlap region.  
The SFR of NGC~4038 is estimated to be $\sim$1~$M_{\solar}$~yr$^{-1}$ 
using the Herschel-PACS data \citep{Klaas10}.  
The molecular gas mass within the central 1~kpc regions of NGC~4038/9 
exceeds that of our Galaxy by a factor of almost 100 \citep{Schulz07}, 
revealing high gas concentration into the two nuclei.  
The extensive amount of molecular gas around the nuclei suggests that 
this galaxy pair may develop into a ULIRG-like star forming galaxy in the future.  
The distribution and characteristics of molecular gas in the Antennae galaxies 
have been investigated using the CO emission \citep[e.g.,][]{Wilson00, Ueda12, Whitmore14} 
and dense gas tracers \citep[i.e., HCN, HCO$^{+}$;][]{Bigiel15, Schirm16}.  
In previous dense gas observations, 
the distribution and properties of molecular gas and star formation rates 
were compared between different regions within the Antennae galaxies.  
In this study, we focus on the nuclear region of NGC~4038, 
which is the northern galaxy of the Antennae galaxies, 
and investigate the distributions of the dense and shocked gas at a 100~pc scale.  
The purposes of this study are to reveal dusty star forming regions 
using dense gas tracers and continuum emission, 
and search for the presence of gas inflows 
using potential molecular line tracers of shocks (SiO, CH$_{3}$OH, HNCO).  
Although it was previously difficult to observe these shock tracers in NGC~4038 due to a lack of sensitivity, 
they are now accessible with the sensitivity achievable with 
the Atacama Large Millimeter/submillimeter Array (ALMA).

\section{Observations and Archival Data}
\subsection{ALMA Multi-line Observations at Band~3}
\begin{figure}[htbp]
	\begin{center}
		\includegraphics[scale=0.6]{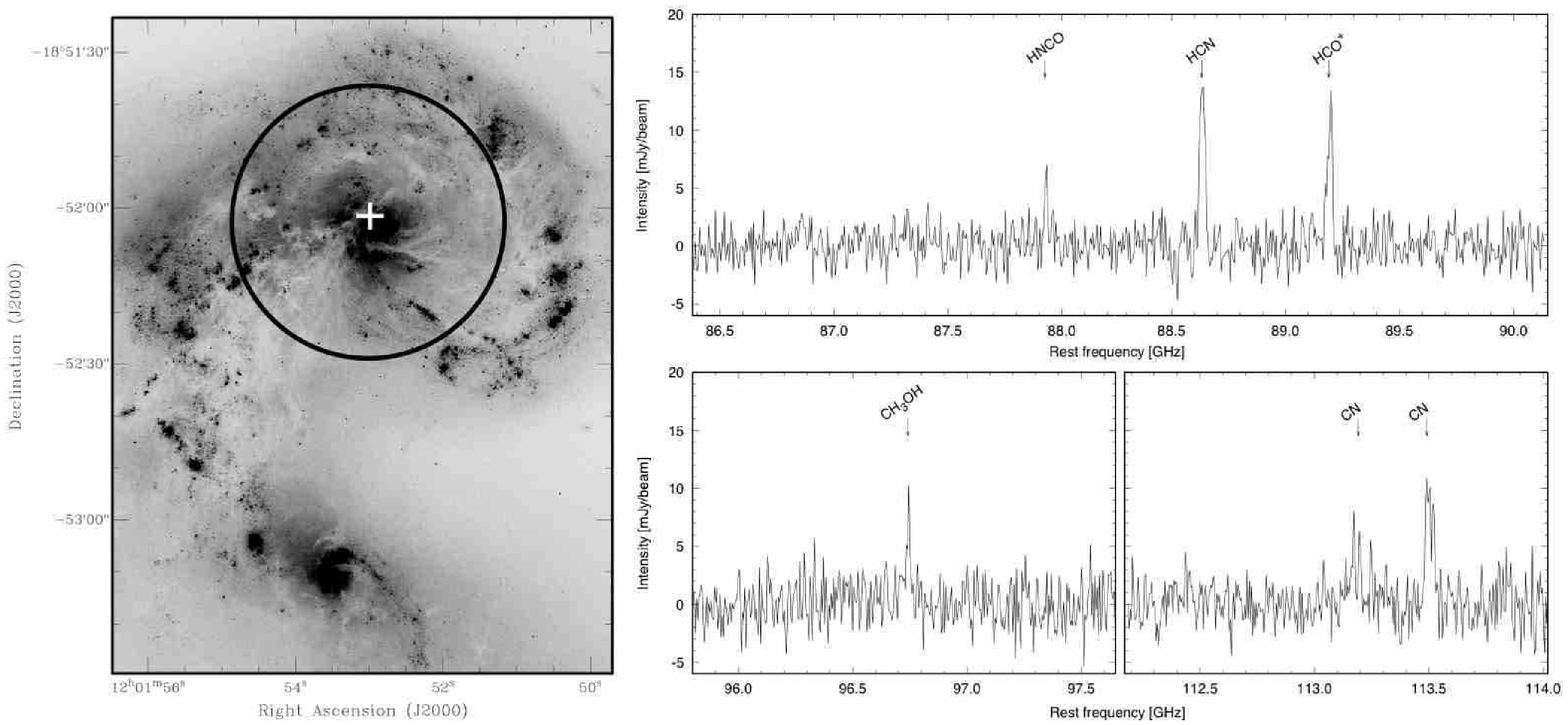}
	\end{center}
	\caption{
	(left) The \textit{HST}/WFC3 F625W image of the Antennae galaxies.  
	The black circle shows the primary beam of the ALMA 12~m antenna.  
	(right) The observed spectra at the strongest CO~(3--2) peak shown by a white cross in the left figure.
	The frequency resolution is 4.88~MHz.
	}
	\label{fig:f1}
\end{figure}

The multi-line observations of NGC~4038 
were carried out using ALMA on July and August, 2014.  
The number of available 12~m antennas ranged 
from 31 to 33 depending on the observing run.  
We used four spectral windows whose frequencies were centered on the lines of 
SiO~(2--1), HCO$^{+}$(1--0), CH$_{3}$OH~(2$_{k}$--1$_{k}$), and CN~(1--0; 1/2--1/2) 
redshifted to the velocity of 1638~km~s$^{-1}$.  
Each spectral window was configured 
with 1.875~GHz bandwidth and 488~kHz frequency resolution.  
The total frequency coverage is 7.47~GHz 
(87.78--91.51~GHz, 97.15--99.02~GHz, and 111.56--113.43~GHz).  
The system temperatures ranged from 40~K to 90~K at the frequencies.  
The data were obtained using a single pointing (Figure~\ref{fig:f1} (left)).  
The primary beam of the 12~m antenna is 53\arcsec--70\arcsec 
depending on the frequency of observations.  
The quasar J1130-1449 was observed for bandpass calibration 
and the quasar J1215-1731 was observed for phase and amplitude calibration.  
Absolute flux calibration was performed using Ceres.  
The uncertainty of absolute flux calibration is 5\% in Band~3 
according to ALMA Cycle~2 Technical Handbook \citep{Lundgren13}.

Data calibration and imaging was carried out 
using the Common Astronomy Software Applications package (CASA; Ver. 4.3.1).  
We used natural weighting of the visibilities for imaging to obtain the highest sensitivity.  
The synthesized beam sizes range between 0\farcs68 and 0\farcs92, 
and the achieved rms noise levels in the velocity resolution of 10~km~s$^{-1}$ 
range between 1.8~mJy~beam$^{-1}$ and 2.4~mJy~beam$^{-1}$ (Table~\ref{tab:t1}).  
We made the 3~mm continuum map using the line free channels.  
The synthesized beam size of the continuum map is 0\farcs95 $\times$ 0\farcs65 
and the rms noise level is 0.05~mJy~beam$^{-1}$.

\subsection{ALMA CO~(3--2) Science Verification Data}
We used ALMA CO~(3--2) Science Verification (SV) data of the Antennae galaxies.  
According to instructions in the CASA guide, 
we applied self-calibration to the calibrated visibility data 
and made the CO~(3--2) map using the CASA (Ver. 4.3.1).  
The synthesized beam size is 1\farcs11 $\times$ 0\farcs65 
by adopting Briggs weighting of the visibilities (robust = 0.5).  
The rms noise level at the velocity resolution of 10~km~s$^{-1}$ is 4.7~mJy~beam$^{-1}$.  
We made the 850~$\mu$m continuum map using the line free channels.  
The synthesized beam size of the continuum map is 1\farcs22 $\times$ 0\farcs73 
by adopting Briggs weighting (robust = 0.5) and the rms noise level is 0.47~mJy~beam$^{-1}$.

\section{Results of the new observations}
\begin{figure}[htbp]
	\begin{center}
		\includegraphics[scale=0.5]{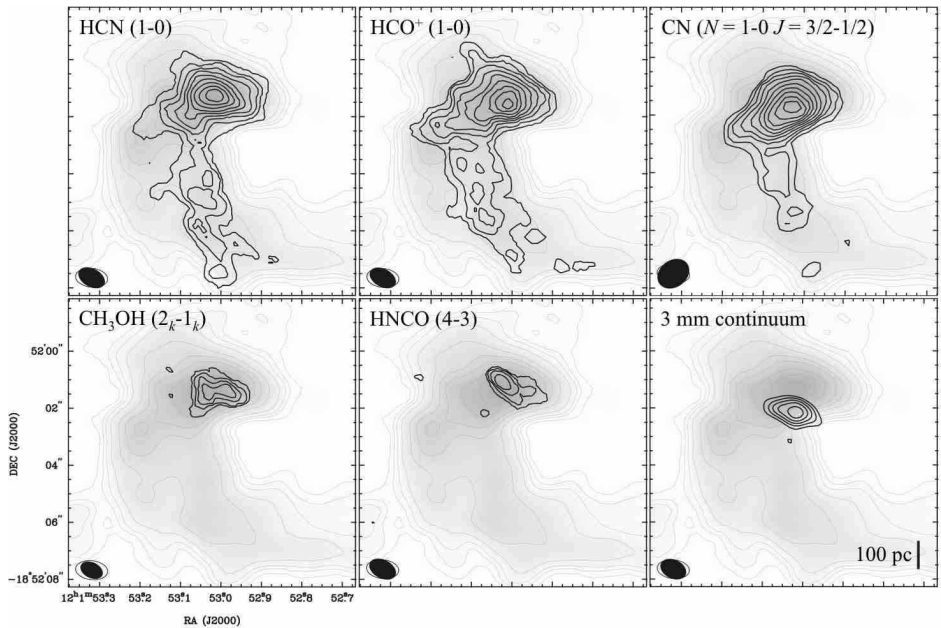}
	\end{center}
	\caption{
	Integrated intensity contour maps of five molecular lines and the 3mm continuum map.
	The background image with gray contours is the CO (3--2) integrated intensity map.
	The filled and open ellipses in the bottom-left corner show the beam sizes of the new map and the CO~(3--2) map, respectively.
	The length of each side corresponds to 10\arcsec ($\sim$1.1~kpc).  
	Contour levels are 120 mJy~beam$^{-1}$ km~s$^{-1}$ $\times$ $n$\, for the HCN map, 
	110 mJy~beam$^{-1}$ km~s$^{-1}$ $\times$ $n$\, for the HCO$^{+}$ map, 
	110 mJy~beam$^{-1}$ km~s$^{-1}$ $\times$ $n$\, for the CN map,
	90 mJy~beam$^{-1}$ km~s$^{-1}$ $\times$ $n$\, for the CH$_{3}$OH map,
	65 mJy~beam$^{-1}$ km~s$^{-1}$ $\times$ $n$\, for the HNCO map ($n$ = 1, 2, 3,...).
	Contour levels are 50 mJy~beam$^{-1}$ $\times$ (3, 4, 5, 6, 7) for the 3~mm continuum map 
	and 2 Jy~beam$^{-1}$ km~s$^{-1}$ $\times$ (1, 2, 3, 5, 10, 15, 20, 30, 40, 50) for the CO map.
	The integrated velocity ranges are 250~km~s$^{-1}$ (1500--1750~km~s$^{-1}$) for the HCN, HCO$^{+}$, CN, and CO maps, 
	100~km~s$^{-1}$ (1600--1700~km~s$^{-1}$) for the CH$_{3}$OH map, 
	and 100~km~s$^{-1}$ (1590--1690~km~s$^{-1}$) for the HNCO map.
	}
	\label{fig:f2}
\end{figure}

The new data contain detections of five molecules 
(CN, HCN, HCO$^{+}$, CH$_{3}$OH, and HNCO), 
including the first detection of three molecules (CN, CH$_{3}$OH, and HNCO) 
in the nuclear region of NGC~4038.  
The observed spectra at the strongest peak of CO~(3--2) are shown in Figure~\ref{fig:f1} (right).  
The peak flux density and integrated line intensity of each line are summarized in Table~\ref{tab:t1}.  
In addition, the 3~mm continuum emission was detected.  
The integrated intensity maps of five molecular lines and the 3~mm continuum map 
overlaid on the CO~(3--2) integrated intensity map are shown in Figure~\ref{fig:f2}.  
The integrated intensity maps were made 
by smoothing and clipping the intensities using the AIPS task, MOMNT.  
The cleaned image cube was smoothed both spatially and in velocity, 
and then this cube was clipped at the 1.5$\sigma$ level per channel.

\subsection{The first detections: CH$_{3}$OH, HNCO, and CN} 
The tracers of shocked gas, CH$_{3}$OH (2$_{k}$--1$_{k}$) 
and HNCO~(4$_{0,4}$--3$_{0,3}$) were detected for the first time.  
The blended CH$_{3}$OH (2$_{k}$--1$_{k}$) line is composed of three transitions, 
(2$_{-1}$--1$_{-1}$) $E$, (2$_{0}$--1$_{0}$) A$^{+}$, and (2$_{0}$--1$_{0}$) $E$.  
The frequency separations between the two neighbor transition lines are less than 3.2~MHz, 
which corresponds to $\lesssim$ 10~km~s$^{-1}$.  
Since the velocity resolution of the channel map that we made is 10 km~s$^{-1}$, 
the emission of these transitions cannot be distinguished.  
Both the CH$_{3}$OH (2$_{k}$--1$_{k}$) and HNCO~(4--3) emission 
are distributed around the strongest CO~(3--2) peak, 
and the spatial distribution of the HNCO~(4--3) is more compact 
than that of the CH$_{3}$OH (2$_{k}$--1$_{k}$) (Figure~\ref{fig:f2}).  
The frequency coverage of our data includes another shock tracer, SiO~(2--1), 
but the SiO~(2--1) emission was not detected.  
The 3$\sigma$ upper limit of the flux density is 5.4 mJy~beam$^{-1}$ 
($\theta_{\rm beam}$ = 1\farcs00 $\times$ 0\farcs65).  
Assuming that the line width of the SiO is the same as the HNCO line width (53~km~s$^{-1}$), 
the 3$\sigma$ upper limit of the integrated intensity is estimated to be 0.30 Jy~km~s$^{-1}$.

Two transitions of CN were also detected for the first time.  
One transition is the CN~(1--0; 3/2--1/2) line which is composed of five hyperfine lines, 
and the other is the CN~(1--0; 1/2--1/2) line which is composed of four hyperfine lines.  
The CN~(1--0; 3/2--1/2)/CN~(1--0; 1/2--1/2) peak intensity ratio is 2.2$\pm$0.2, 
which is consistent with the statistical value within the uncertainties, 
suggesting that the CN~(1--0; 1/2--1/2) line is optically thin.  
Since the CN~(1--0; 1/2--1/2) emission is very faint, 
we use the CN~(1--0; 3/2--1/2) (hereafter CN~(1--0)) map in the following discussion.

\subsection{The HCN and HCO$^{+}$ emission}
The HCN~(1--0) and HCO$^{+}$(1--0) emission was clearly detected  in the central 1~kpc region.  
Both emission are concentrated in a region associated with the strongest CO~(3--2) peak, 
but the HCN and HCO$^{+}$ peaks are offset by 0\farcs4.  
The HCN and HCO$^{+}$ integrated intensities are 
82\% and 89\% of previous measurements with ALMA \citep{Schirm16}, respectively.  
These differences are due to the limited maximum recoverable scale (MRS).  
The MRS of our data is 1.3~kpc, which is 2.7 times smaller 
than the MRS of the previous data \citep{Schirm16}.  
We estimate the HCN~(1--0)/HCO$^{+}$(1--0) line ratio
using the integrated intensity maps clipped at 2$\sigma$ levels.  
We ignore a slight (0\farcs01) difference between the beam sizes of the HCN and HCO$^{+}$ maps.  
The average luminosity ratio of $L_{\rm HCN}$/$L_{\rm HCO^{+}}$ is 1.2 $\pm$ 0.9.  
The standard deviation quoted reflects the distribution of the HCN/HCO$^{+}$ luminosity ratios 
(the uncertainty based on the calibration uncertainties is 1.2 $\pm$ 0.1).  
This average ratio is 1.2 times higher than the previous measurement 
\citep[$L_{\rm HCN}$/$L_{\rm HCO^{+}}$ = 1.019 $\pm$ 0.008;][]{Schirm16}.  
This difference could be affected by the difference of the MRS and clipping.  
The variation of the line ratios may show differences of abundances and excitation, 
and additional data are required to confirm the causes of the various line ratios.

\subsection{The 3~mm Continuum emission}
The total flux of the 3~mm continuum estimated 
by integrating the extended emission is 0.59 $\pm$ 0.03~mJy.  
The 3~mm continuum peak is located 
$\sim$1\arcsec (100~pc) south of the strongest CO~(3--2) peak 
and corresponds to the 6~cm and 4~cm continuum sources.  
In order to estimate the radio spectral index with beam-matched radio maps, 
we reimaged the 6~cm and 4~cm continuum emission 
using the VLA archival calibrated data (PI: A. Pedlar) obtained with A and B configurations.  
We clipped the visibilities so that the VLA and ALMA data 
have the same shortest uv range (9.7~k$\lambda$) 
and convolved both data to 1\farcs2 angular resolution.  
The peak flux of the 6~cm and 4~cm continuum are 
0.94 $\pm$ 0.07~mJy and 0.84 $\pm$ 0.09~mJy, respectively.  
The radio spectral index is estimated to be $\alpha$ = -0.18 $\pm$ 0.34, 
indicating that the 6~cm and 4~cm continuum emission is dominated 
by free-free emission from ionized gas.  
Using this radio spectral index, 
the continuum flux at 3~mm is estimated to be 0.55 $\pm$0.24~mJy.  
This is consistent with the observed value within the error.  
Thus the 3~mm continuum emission is dominated by free-free emission.

\section{Discussion}
The following discussion in \S4.1--\S4.3 are based on maps 
which were created with visibilities clipped to have similar shortest uv range 
(uv distance = 12-14 k$\lambda$), and convolved to 1\farcs2 angular resolution.  
The maps are not sensitive to extended ($>$1.0kpc) structures.

\subsection{Photo-Dominated Region}
\begin{figure}[htbp]
	\begin{center}
		\includegraphics[scale=0.5]{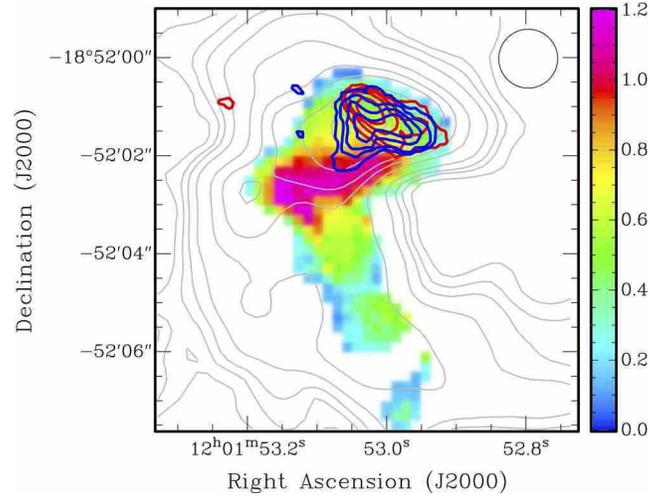}
	\end{center}
	\caption{
	The color-scale image is the CN~(1--0; 3/2--1/2)/HCN~(1--0) integrated intensity ratio map
	with a resolution of 1\farcs2 (shown in the upper-right corner).  
	The blue, red, and gray contours show the CH$_{3}$OH~(2$_{k}$--1$_{k}$), 
	HNCO~(4--3), and CO~(3--2) integrated intensity maps, respectively.  
	The contour levels are the same as Figure~\ref{fig:f2}. 
	}
	\label{fig:f3}
\end{figure}

The CN molecule is another tracer of dense gas, 
with a lower (by a factor of five) critical density than the HCN \citep{Aalto02}, 
and the CN/HCN line ratio has been considered 
to be a tracer of dense Photo-Dominated Regions \citep[PDRs; e.g.,][]{Fuente93, Boger05}.  
It is known that the CN/HCN line ratio increases with the strength of interstellar UV radiation field.  
We made the CN~(1--0)/HCN~(1--0) line ratio map (Figure~\ref{fig:f3}) 
using the convolved CN~(1--0) and HCN~(1--0) integrated intensity maps clipped at the 2$\sigma$ levels.  
Although the average ratio (CN/HCN = 0.7 $\pm$ 0.3) is lower than the global CN~(1--0)/HCN~(1--0) line ratios 
measured in typical starburst galaxies \citep[e.g., Arp~220 and NGC~253;][]{Baan08}, 
we find a gradient in the CN/HCN line ratio.  
The enhanced CN/HCN line ratios ($\simeq$ 0.8--1.2) are seen 
in the southern region of the strongest CO~(3--2) peak (Figure~\ref{fig:f3}), 
where the 3~mm continuum peak is also located.  
Although the resolution of the \textit{Spitzer}/MIPS 24~$\mu$m image is 6\arcsec, 
which is five times lower than the resolutions of the convolved CN and HCN maps, 
the 24 $\mu$m peak is located 200~pc south of the strongest CO~(3--2) peak, 
and roughly corresponds to the region associated with high CN/HCN line ratios.  
Thus, the enhanced CN/HCN line ratios suggest the presence of PDRs.

It is also known that the C$_{2}$H emission is enhanced in PDRs \citep[e.g.,][]{Fuente93}.  
Although the frequency coverage of our data included 
the set of the C$_{2}$H~(1--0) lines whose rest frequencies are around 87.4~GHz,
the emission was not detected.
The 3~$\sigma$ upper limit of the flux density is 5.4~mJy~beam$^{-1}$.
From spectra obtained by \citet{Fuente05}, 
the C$_{2}$H~(1--0)/HCN~(1--0) peak intensity ratios are estimated to be 0.2--0.4 
in the central  650~pc disk in M82 where PDRs form.
Assuming the same C$_{2}$H/HCN line ratios, 
the C$_{2}$H peak flux density is expected to be $\leq$3.0~mJy~beam$^{-1}$ 
for regions in NGC~4038 where the CN/HCN line ratios are larger than 0.8.    
This expected peak flux density is smaller than the 3$\sigma$ upper limit.  
New maps with higher sensitivity are required 
to investigate the properties of the C$_{2}$H molecules 
in the central region of NGC~4038.

\subsection{Dense gas fraction}
We investigate the dense gas fraction in the central region of NGC~4038.  
\citet{Bigiel15} estimate the dense gas fraction in the Antennae galaxies 
using low-resolution ($\theta_{\rm beam}$ = 6\farcs5) HCN~(1--0) and CO~(1--0) maps.  
The dense gas fraction is estimated to be $\sim$11~\% 
in the central $\sim$700~pc region of NGC~4038 \citep[Region ID = 1;][]{Bigiel15}.  
\citet{Schirm16} estimate the HCN~(1--0)/CO~(1--0) line ratio 
using the HCN~(1--0) and CO~(1--0) maps with angular resolutions of $\sim$4\arcsec, 
though they do not convert the line ratio into the dense gas fraction.  
Since a high-resolution ($\leq$4\arcsec) CO~(1--0) map is not currently available, 
we estimate the dense gas fraction from the HCN~(1--0)/CO~(3--2) line ratio, 
assuming a single CO~(3--2)/CO~(1--0) line ratio.  
Although CO~(3--2) is not the best tracer of diffuse molecular gas, 
the critical density of CO~(3--2) is two orders of magnitude lower than that of the HCN~(1--0).  
The resolution of the CO~(3--2) SV data is comparable to that of the new HCN~(1--0) data, 
and these data allow us to compare the gas distributions at a 100~pc scale.  
The luminosity ratios of $L_{\rm HCN}$/$L_{\rm CO~(3-2)}$ range from 0.01 to 0.34, 
suggesting that the dense gas fraction vary within the central 1~kpc.  
One peak of the line ratio is located on the northwest side of the strongest HCN~(1--0) peak, 
and the other peaks are found at the south end of the HCN distribution 
and at the HCN peak located in the middle of the HCN distribution.  
We convert the HCN~(1--0)/CO~(3--2) line ratios into HCN~(1--0)/CO~(1--0) line ratios, 
assuming the CO~(3--2)/(1--0) line ratio of 0.6, 
which was measured using the CO~(1--0) and CO~(3--2) maps of NGC~4038 
with $\sim$4\farcs6 resolution \citep{Ueda12}.  
The average luminosity ratio of $L_{\rm HCN}$/$L_{\rm CO~(1-0)}$ is 0.08 $\pm$ 0.03.  
This is consistent with a previous study by \citet{Schirm16} 
using maps with a resolution of $\sim$4\arcsec.

We estimate the dense gas mass fraction ($M_{\rm HCN}/M_{\rm H_{2}}$) 
using the same conversion factors as the previous study \citep{Bigiel15} 
in order to estimate the dense gas mass fraction under the same assumptions.  
We use a Galactic CO-to-H$_{2}$ mass conversion factor 
$\alpha_{\rm CO}$ = 4.35 $M_{\solar}$ (K~km~s$^{-1}$~pc$^{2}$)$^{-1}$ \citep[e.g.,][]{Bolatto13} 
and the HCN-to-dense gas mass conversion factor 
$\alpha_{\rm HCN}$ = 10 $M_{\solar}$ (K~km~s$^{-1}$~pc$^{2}$)$^{-1}$ \citep{Gao04}.  
The average dense gas fraction is estimated to be 17 \% $\pm$ 8\%, 
which is larger than previously reported \citep[$\sim$11\% measured in ID=1 by][]{Bigiel15}.  
This is expected, since the CO~(3--2) emission comes from more compact gas cores 
than extended gas components traced in the CO~(1--0).  
We note that our estimation has a large uncertainty 
because the dense gas fraction changes 
depending on the assumed CO~(3--2)/CO~(1--0) line ratio.  
Assuming the CO~(3--2)/CO~(1--0) line ratio of 0.7 and 0.5, 
the average dense gas fraction changes to $\sim$20\% and $\sim$14\%, respectively.  
Furthermore, \citet{Zhu03} propose a conversion factor for the nucleus of NGC~4038 
of 2.3 $\times$ 10$^{19}$ cm$^{-2}$ (K~km~s$^{-1}$)$^{-1}$, 
which corresponds to $\alpha_{\rm CO} \sim$ 0.4.  
This is approximately an order of magnitude lower than the Galactic value, 
and it is also lower than the conversion factor in gas-rich galaxies at high redshift 
and local luminous infrared galaxies \citep[$\alpha_{\rm CO}$ = 0.6--0.8; e.g.,][]{Papadopoulos12}.  
Adapting this conversion factor, the average dense gas fraction exceeds 100\%.  
While this is not plausible, it shows that the dense gas fraction 
can change significantly depending on the mass conversion factor.  
Also, the HCN-to-dense gas mass conversion factor may vary and affect the dense gas fraction.  
These two conversion factors should be investigated in future studies.

\subsection{Systematic distributions of different molecular species and continuum emission}
\begin{figure}[htbp]
	\begin{center}
		\includegraphics[scale=0.95]{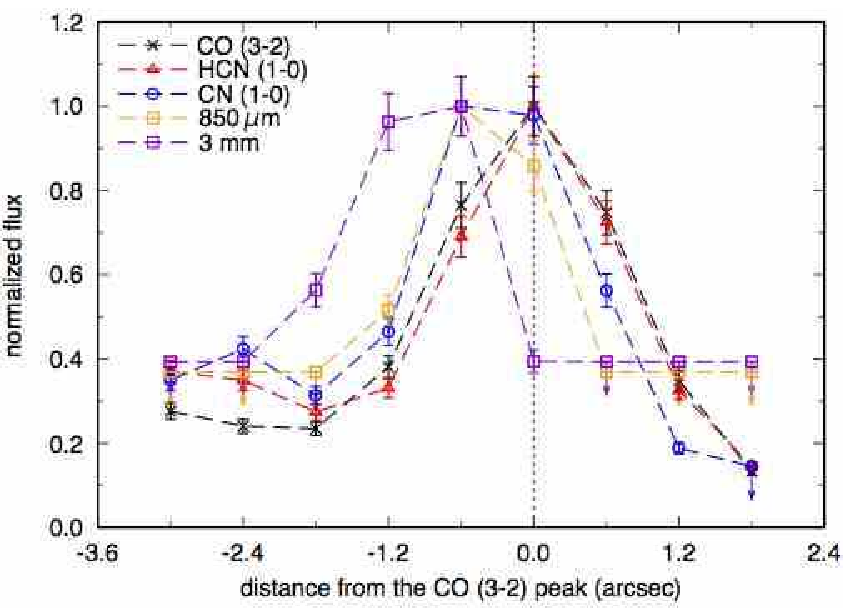}
		\includegraphics[scale=0.95]{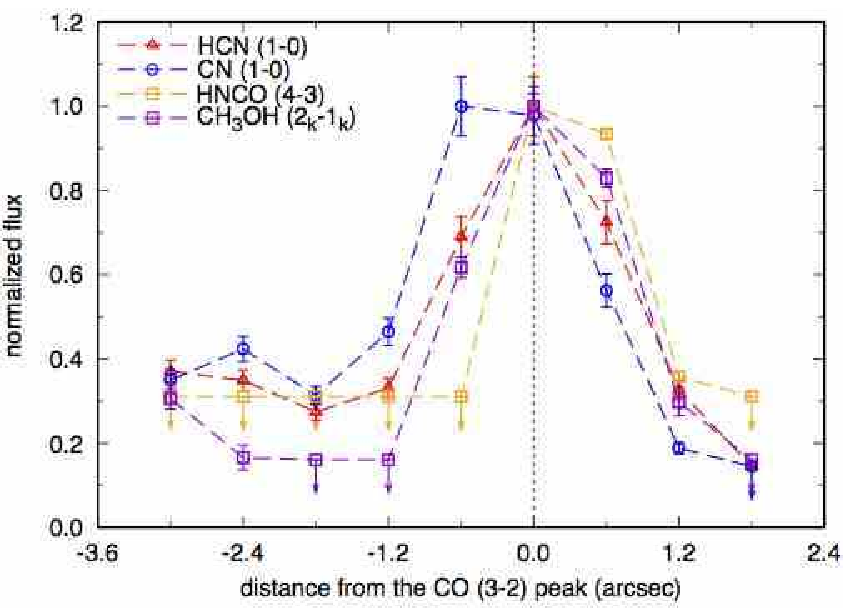}
 	\end{center}
	 \caption{
	 (left) Flux profiles of the CO~(3--2), HCN~(1--0), CN~(1--0), 850~$\mu$m continuum, and 3~mm continuum emission 
	 as a function of the projected distance from the strongest CO~(3--2) peak.
	 The positive and negative values correspond to the north and south directions.
	 (right) Same as the left figure, but for the HCN~(1--0), CN~(1--0), HNCO~(4--3), and CH$_{3}$OH~(2$_{k}$--1$_{k}$) lines.
	}
 	\label{fig:f4}
\end{figure}

The flux profiles as a function of the projected distance 
from the strongest CO~(3--2) peak are shown in Figure~\ref{fig:f4} (left).  
We measured the integrated intensities of the CN~(1--0), HCN~(1--0), and CO~(3--2), 
and the 3~mm and 850~$\mu$m continuum flux 
in nine points arranged in the north-south direction, 
which are roughly located along strong CO~(3--2) emission.  
The separation between the measurement points is 0\farcs6, 
which is a half of the convolved beam size.  
The HCN profile is very similar to the CO profile.  
The CN profile is slightly different from the HCN and CO profiles 
and similar to the 850~$\mu$m continuum profile.  
The peak of the CN integrated intensity shifts toward the south 
compared to the HCN and CO peaks.  
The 3~mm continuum profile shifts further toward the south.  
These comparisons reveal the systematic differences 
in distribution of the different molecular species and the continuum emission, 
and in particular, the softening of the interstellar radiation field 
(as traced by the CN/HCN ratio) as a function of distance from the star forming region.  
Such systematic differences are found 
in the supergiant H\emissiontype{II} region in M33 \citep{Miura10}.  
A star forming region identified by the free-free emission at 3~mm 
is located $\sim$100~pc south of the CO peak.  
The formation of high mass stars is currently taking place 
at the 850~$\mu$m emission peak, and it is located $\sim$50~pc south from the CO peak.  
It is likely that PDRs are forming in these star forming regions 
because the CN/HCN line ratios are enhanced (CN/HCN $\simeq$ 0.8--1.2).  
The HCN~(1--0)/CO~(1--0) luminosity ratios 
estimated from the HCN~(1--0) and CO~(3--2) maps (\S~4.1) 
are $>$0.09 around the CO peak, 
which are comparable to the HCN~(1--0)/CO~(1--0) line ratios ($\sim$0.1) 
in molecular clouds in the nuclear starburst region of NGC~253 \citep{Leroy15}.  
This implies a large concentration of dense gas (the dense gas fraction $>$20\%) 
around the CO peak in NGC~4038.  
Furthermore, the molecular gas mass within a 0\farcs6 ($\sim$60~pc) radius of the CO peak 
is estimated to be 1.1 $\times$ 10$^{8}$ $M_{\solar}$, 
using the CO~(3--2)/CO~(1--0) line ratio of 0.6 \citep{Ueda12} 
and $\alpha_{\rm CO}$ = 4.35 $M_{\solar}$~(K~km~s$^{-1}$~pc$^{2}$)$^{-1}$ \citep{Bolatto13}.  
The large molecular gas mass and high dense gas fraction 
may signify intense star formation around the CO peak.

\subsection{Different distributions of two shock tracers}
The CH$_{3}$OH and HNCO emission in galaxies has been used 
to trace large-scale molecular shocks \citep[e.g.,][]{Garcia-Burillo01, Meier05}.  
Using a CH$_{3}$OH map convolved to the beam size of the HNCO map, 
the CH$_{3}$OH~(2$_{k}$--1$_{k}$)/HNCO~(4--3) peak intensity ratio 
is 1.7$\pm$0.1 in the CH$_{3}$OH emission peak.  
This is similar to the CH$_{3}$OH~(2--1)/HNCO~(4--3) ratios measured in other galaxies.  
The CH$_{3}$OH~(2--1)/HNCO~(4--3) ratios are 
1.4 for the northern arm of IC~342 \citep{Meier05} and 
1.5 for the circum-nuclear disk of NGC~1068 \citep{Garcia-Burillo10}.  
On the other hand, the peak intensity ratio is 0.79$\pm$0.06 in the HNCO emission peak, 
which is about half of the ratio in the CH$_{3}$OH emission peak.  
Furthermore, the distribution of the HNCO emission is different 
from the CH$_{3}$OH distribution (Figure~\ref{fig:f3}).  
This could be due to the stellar UV field.  
The HNCO distribution can be affected by strong UV field 
because the HNCO molecule is easily photo-dissociated \citep{Martin08}.  
Another possibility is that the CH$_{3}$OH is enhanced by UV field 
because of desorption into gas-phase from dust by UV photons \citep{Guzman13}.  
As shown in Figure~\ref{fig:f3}, the CH$_{3}$OH distribution is more extended than the HNCO, 
and the CH$_{3}$OH emission is detected in a region associated with high ($\geq$0.8) CN/HCN line ratios.  
The flux profiles (Figure~\ref{fig:f4} (right)) also show 
that the HNCO emission is diminished around the CN peak compared to the CH$_{3}$OH emission.  
Since the CN/HCN line ratio is a tracer of dense PDR, 
the different distributions of CH$_{3}$OH and HNCO could be caused by the stellar UV field.   

\subsection{Possibilities of the shock origin}
\begin{figure}[htbp]
	\begin{center}
		\includegraphics[scale=0.65]{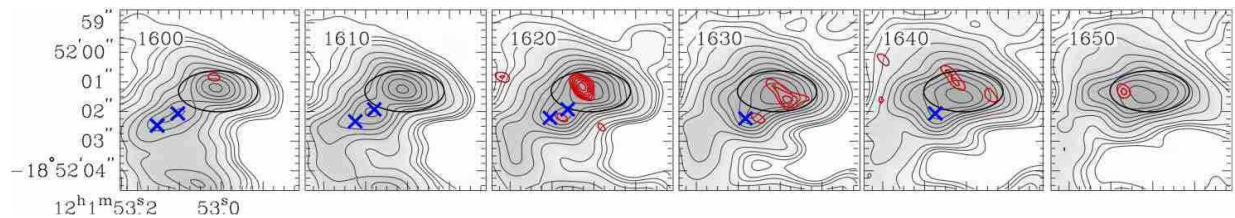}
	\end{center}
	\caption{
	The gray-scale image with black contours is the CO~(3--2) channel map 
	with the velocity resolution of 10~km~s$^{-1}$ in the central $\sim$300~pc region.  
	The contour levels are 23.5~mJy~beam$^{-1}$ $\times$ (1, 2, 3, 5, 10, 15, 20, 25, 30, 35, 40, 45, 50).  
	The red contours (1.8~mJy~beam$^{-1}$ $\times$ (3.0, 3.5, 4.0, 4.5, 5.0, 5.5)) show the HNCO~(4--3) channel map. 
	The black ellipse roughly shows the location of Complex~1 and the blue cross symbols show the inflowing gas (see \S4.5).     
	}
	\label{fig:f5}
\end{figure}

One possibility of the shock origin is collisions 
between inflowing gas and a massive molecular complex.  
We applied the automatic clump identification algorithm Clumpfind \citep{Williams94} 
to the ALMA CO~(3--2) SV data, 
using the recommended 2$\sigma$ threshold to identify robust molecular clumps.  
The most massive molecular complex (Complex~1) 
is identified in a region associated with the strongest CO~(3--2) peak (Figure~\ref{fig:f5}).  
The radius and systemic velocity of Complex~1 
are 144 $\pm$ 30~pc and 1620~km~s$^{-1}$, respectively.  
The molecular gas mass of Complex~1 is estimated to be $3.2 \times 10^{8}~M_{\solar}$, 
using the CO~(3--2)/CO~(1--0) line ratio of 0.6 \citep{Ueda12} 
and $\alpha_{\rm CO}$ = 4.35 $M_{\solar}$~(K~km~s$^{-1}$~pc$^{2}$)$^{-1}$ \citep{Bolatto13}.  
The molecular gas indicated by blue cross symbols in Figure~\ref{fig:f5} 
moves toward Complex~1 and merges with Complex~1 
as the velocity goes from 1600~km~s$^{-1}$ to 1650~km~s$^{-1}$.  
This component appears to inflow toward Complex~1 
because the line-of-sight velocity of the gas component 
is different from a velocity expected from the galactic rotation of NGC~4038.
According to the velocity fields of atomic gas \citep{Hibbard01} and ionized gas \citep{Amram92}, 
NGC~4038 has a single trailing arm extending toward the southwest.
The southwest part of this galaxy is located on the front side, 
and the northeast part is located on the rear side.  
Based on this geometry, the line-of-sight velocities of molecular gas distributed along the arm 
should decrease as closer to the center (i.e., Complex~1).  
However, the line-of-sight velocities of the gas indicated by blue cross symbols in Figure~\ref{fig:f5} 
increase as closer to Complex~1.
We thus suggest that the gas component is inflowing toward Complex~1.  
We note that we cannot completely reject other possibilities (e.g., outflow)
because the structure and velocity field in the central region 
are significantly disturbed and complicated due to the dynamical interaction.
The shocked gas traced in the CH$_{3}$OH and HNCO emission is seen around a region 
where the inflowing gas component connects with Complex~1, 
and the line-of-sight velocities of the shocked gas are consistent 
with those of the inflowing gas component and Complex~1.  
Therefore, we suggest that the CH$_{3}$OH and HNCO emission traces molecular shocks 
caused by collisions between the inflowing gas and the central massive molecular complex.  

Another possibility of the shock origin is outflows induced by star formation, 
as often demonstrated in Galactic sources.  
A typical size of shock regions created by star formation is $\sim$0.1--1~pc \citep[e.g., L1157;][]{Gueth98}, 
but the shocked gas is distributed across a few 100~pc in the nuclear region of NGC~4038.  
Such a large-scale shock can be caused by a kinematic factor rather than by star formation.  
It is likely that the emission of the shock tracers comes from multiple sites of cloud-cloud collisions.  
In addition, we estimate the number density of outflows 
which would be required to reproduce the observed CH$_{3}$OH intensity in the nuclear region of NGC~4038, 
using the CH$_{3}$OH luminosity measured in the blue lobe of L1157.  
The same discussion has been conducted using the SiO emission by \citet{Usero06}.
The mean CH$_{3}$OH intensity in NGC~4038 is 5.55~mJy~km~s$^{-1}$, 
which corresponds to 1.60~K~km~s$^{-1}$.  
According to \citet{Usero06}, the CH$_{3}$OH luminosity is 0.50~K~km~s$^{-1}$~pc$^{2}$ in the blue lobe of L1157.
Thus the required number density of outflows is estimated to be 3.2 outflows pc$^{-2}$. 
Although star-forming regions with a few outflows pc$^{-2}$ are found \citep[e.g.,][]{Williams03}, 
it is unlikely that such star-forming regions is forming across a few 100~pc$^{2}$.
This is the same conclusion as the previous study towards IC~342 \citep{Usero06}.

Finally, we discuss shock strength using the CH$_{3}$OH/SiO and HNCO/SiO line ratios.  
The CH$_{3}$OH/SiO line ratio is used as an indicator of shock strength 
because the CH$_{3}$OH molecules could be destroyed 
by relatively slow shocks ($v_{\rm shock}$ $>$ 10--15~km~s$^{-1}$) 
and the dissociation of the SiO molecules would require fast shocks 
($v_{\rm shock}$ $>$ 50--60~km~s$^{-1}$) \citep{Usero06}.  
The HNCO/SiO line ratio can be also an indicator of shock strength 
because the HNCO molecules could be sublimated by weak shocks \citep{Rodriguez-Fernabdez10, Meier15}.  
We estimate the lower limits of the CH$_{3}$OH/SiO and HNCO/SiO line ratios in the nuclear region of NGC~4038, 
using the 3$\sigma$ upper limit of the SiO flux density (5.4~mJy~beam$^{-1}$).  
The lower limits of the CH$_{3}$OH/SiO and HNCO/SiO peak intensity ratios are 2.0 and 1.9, respectively.  
These are comparable to the line ratios measured in
in both the young bipolar outflow L1157 and the circum-nuclear disk of NGC~1068 \citep{Garcia-Burillo10}.  
This suggests that fast shocks may have occurred in the nuclear region of NGC~4038.  
However, if the SiO emission is much weaker than the 3$\sigma$ upper limit, 
the line ratios become larger, indicating that a possibility for the presence of fast shocks is reduced.  
High-sensitivity SiO measurements are required 
to draw a conclusion about the shock strength in the nuclear region of NGC 4038.

\section{Summary}
We have conducted multi-line observations ($\lambda$ = 3~mm) of NGC~4038 using ALMA.  
Five molecules (CN, HCN, HCO$^{+}$, CH$_{3}$OH, and HNCO) were detected in the central $\leq$1~kpc region, 
and three (CN, CH$_{3}$OH, and HNCO) of them was detected for the first time in this region.  

The flux profiles as a function of the distance from the CO~(3--2) peak reveal 
the systematic differences in distributions of the different molecular species and continuum emission.  
A star forming region identified by the free-free emission (the 3~mm continuum emission) 
and  the most active site of dusty star forming region (the 850~$\mu$m continuum emission) 
are located $\sim$100~pc and $\sim$50~pc, respectively, south of the HCN~(1--0) and CO~(3--2) peaks.  
The CN~(1--0)/HCN~(1--0) line ratios are enhanced (CN/HCN $\simeq$ 0.8--1.2) 
in these star forming regions, suggesting the presence of PDRs.  
Assuming the CO~(3--2)/CO~(1--0) line ratio of 0.6, 
the average HCN~(1--0)/CO~(1--0) luminosity ratio is estimated to be 0.08 $\pm$ 0.03.  
This corresponds to the dense gas fraction of 17\% $\pm$ 8\%.  
The large molecular gas mass (10$^{8}$~$M_{\solar}$) and high dense gas fraction ($>$20\%) 
are found around the CO peak, which suggests potential star formation there.

The detection of the shocked gas traced in the CH$_{3}$OH and HNCO emission 
is evidence for sub-kpc scale molecular shocks.  
The distribution of the HNCO is slightly different from the CH$_{3}$OH distribution.  
The CH$_{3}$OH emission is detected in a region associated with high ($\geq$0.8) CN/HCN line ratios. 
Since the CN/HCN line ratio is a tracer of PDR, 
the different distributions of CH$_{3}$OH and HNCO could be caused by the stellar UV field.  
Comparing the distributions of the shocked gas and the CO~(3--2) emission, 
we suggest that the molecular shocks could be caused by collision 
between inflowing gas and the central massive molecular complex.

\begin{ack}
This paper has made use of the following ALMA data: ADS/JAO.ALMA\#2012.0.01000.S 
and ALMA Science Verification data: ADS/JAO.ALMA\#2011.0.00003.SV.
ALMA is a partnership of the ESO (representing its member states), 
NSF (USA), and NINS (Japan), together with the NRC (Canada), 
NSC, and ASIAA (Taiwan), in cooperation with the Republic of Chile.  
The Joint ALMA Observatory is operated by the ESO, AUI/NRAO, and NAOJ.  

One image presented in this paper were obtained from the Mikulski Archive for Space Telescopes (MAST).  
STScI is operated by the Association of Universities for Research in Astronomy, Inc., under NASA contract NAS5-26555.  
Support for MAST for non-HST data is provided by the NASA Office of Space Science 
via grant NNX09AF08G and by other grants and contracts.

Data analysis were in part carried out on common use data analysis computer system 
at the Astronomy Data Center, ADC, of the National Astronomical Observatory of Japan.

J.U. was supported by the ALMA Japan Research Grant of NAOJ Chile Observatory, 
NAOJ-ALMA-0058.
S.O. and T.S. are financially supported by a Research Fellowship
from the Japan Society for the Promotion of Science for Young Scientists.
\end{ack}

\onecolumn

\begin{table}[htbp]
\tbl{Properties of ALMA data for the detected molecular lines}{%
\begin{tabular}{lcrcrccc}
\hline
\multicolumn{1}{c}{Molecule}&\multicolumn{1}{c}{Transition}&\multicolumn{1}{c}{Frequency}&\multicolumn{1}{c}{Beam Size}
&\multicolumn{1}{c}{P.A.}&\multicolumn{1}{c}{rms\footnotemark[$*$]}&\multicolumn{1}{c}{Peak flux density}&\multicolumn{1}{c}{Integrated intensity}\\
&&\multicolumn{1}{c}{[GHz]}&\multicolumn{1}{c}{[arcsec]}
&\multicolumn{1}{c}{[degree]}&\multicolumn{1}{c}{[mJy Beam$^{-1}$]}&\multicolumn{1}{c}{[mJy beam$^{-1}$]}&\multicolumn{1}{c}{[Jy km s$^{-1}$]}\\
\hline
HNCO & 4$_{0,4}$--3$_{0,3}$ & 87.925 & 1.00 $\times$ 0.65 & 62.8 & 1.8 & 10.5 $\pm$ 0.5 & 0.66 $\pm$ 0.03\\
HCN & $J$ = 1--0 & 88.632 & 1.00 $\times$ 0.64 & 61.7 & 1.9 & 15.4 $\pm$ 0.8 & 11.7 $\pm$ 0.6\\
HCO$^{+}$ & $J$ = 1--0 & 89.189 & 0.99 $\times$ 0.63 & 61.8 & 1.8 & 14.5 $\pm$ 0.7 & 12.8 $\pm$ 0.6\\
CH$_{3}$OH & 2$_{k}$--1$_{k}$ & 96.741 & 0.83 $\times$ 0.55 & 64.4 & 2.3 & 10.8 $\pm$ 0.5 & 1.74 $\pm$ 0.09\\
CN & $N$ = 1--0 $J$ = 1/2--1/2 & 113.191 & 1.19 $\times$ 0.93 & 126.4 & 2.0 & 8.4 $\pm$ 0.4 & 1.0 $\pm$ 0.1\\
CN & $N$ = 1--0 $J$ = 3/2--1/2 & 113.491 & 1.19 $\times$ 0.93 & 126.4 & 2.1 & 18.5 $\pm$ 0.9 & 5.7 $\pm$ 0.3\\
\hline
\end{tabular}}
\label{tab:t1}
\begin{tabnote}
\hangindent6pt\noindent
\hbox to6pt{\footnotemark[$*$]\hss}\unskip%
The noise level in the velocity resolution of 10~km~s$^{-1}$.
\end{tabnote}
\end{table}

\end{document}